\documentclass[letterpaper,twocolumn,10pt]{article}
\usepackage{usenix2020_SOUPS}
\usepackage{tikz}
\usepackage{amsmath}
\usepackage{cleveref}

\microtypecontext{spacing=nonfrench}
\begin{document}

\date{}
\title{Towards Secure and Usable Authentication for\\ Augmented and Virtual Reality Head-Mounted Displays
}

\def\plainauthor{Reyhan Duezguen, Peter Mayer, Sanchari Das, Melanie Volkamer}
\author{
\rm{Reyhan Duezguen$^{\ast}$, Peter Mayer$^{\ast}$, Sanchari Das$^{\dagger}$, Melanie Volkamer$^{\ast}$}\\
$^{\ast}$ SECUSO - Security, Usability, Society, Karlsruhe Institute of Technology\\
$^{\dagger}$ University of Denver, Indiana University Bloomington \\
$^{\ast}$ firstname.lastname@kit.edu, 
$^{\dagger}$  sanchari.das@du.edu
}
\maketitle
\thecopyright

\begin{abstract}
Immersive technologies, including augmented and virtual reality (AR \& VR) devices, have enhanced digital communication along with a considerable increase in digital threats. Thus, authentication becomes critical in AR \& VR technology, particularly in shared spaces. In this paper, we propose applying the ZeTA protocol that allows secure authentication even in shared spaces for the AR \& VR context. We explain how it can be used with the available interaction methods provided by Head-Mounted Displays. In future work, our research goal is to evaluate different designs of ZeTA (e.g., interaction modes) concerning their usability and users' risk perception regarding their security - while using a cross-cultural approach. 
\end{abstract}

\section{Introduction}
New-age technologies help to connect people despite geographical constraints. However, such technological evolution brings new risks. Augmented and virtual reality (AR \& VR) are such technologies that have expanded considerably and are projected to reach \$114 billion and \$65 billion, respectively, by 2021~\cite{ABI2017}. AR \& VR systems like the Oculus and Google Glass increasingly promise to provide social activities like interactive gaming, virtual shopping, or attending virtual meetings~\cite{roberts2014visualization}. Many of these activities happen in so-called shared spaces, i.e., places not strictly public, but where multiple people are present at the same time~\cite{gugenheimer2019challenges}. However, these technologies also introduce new security challenges in AR \& VR~\cite{happa2019cyber}, including authentication challenges. Nowadays, authentication on AR \& VR systems is neglected or carried out on the smartphone or PC~\cite{chan2015glass}. Yet, if authentication is required during a VR experience, e.g., paying for a product or entering a virtual conference, the user must take off the Head-Mounted Display (HMD), interrupting the virtual experience. Such challenges motivated our research direction to implement more secure and usable authentication strategies for AR \& VR devices.

A naive approach using voice recognition technology of the HMD as an authentication strategy might put users at serious security risks, especially in public and shared spaces. Another method could be to use the available sensors for biometric authentication, e.g., gait recognition~\cite{gafurov2006biometric}. Such authentication schemes are designed for continuous authentication. The goal of our research is to focus on authenticating services when needed. Additionally, biometric-based approaches would also hamper authenticating with someone else's HMD (as it would first need to be trained) and may have several privacy concerns. Thus, what is needed is a secure (especially in shared spaces) and usable authentication scheme, which only uses the sensors of the HMDs while being privacy-preserving. 

Therefore, we are 
    {\it \textbf {proposing a shoulder-surfing resistant authentication scheme that relies only on the equipment of the AR \& VR HMDs.} }
    
\begin{figure*}[htp!]
    \centering
    \includegraphics[width=0.8\linewidth]{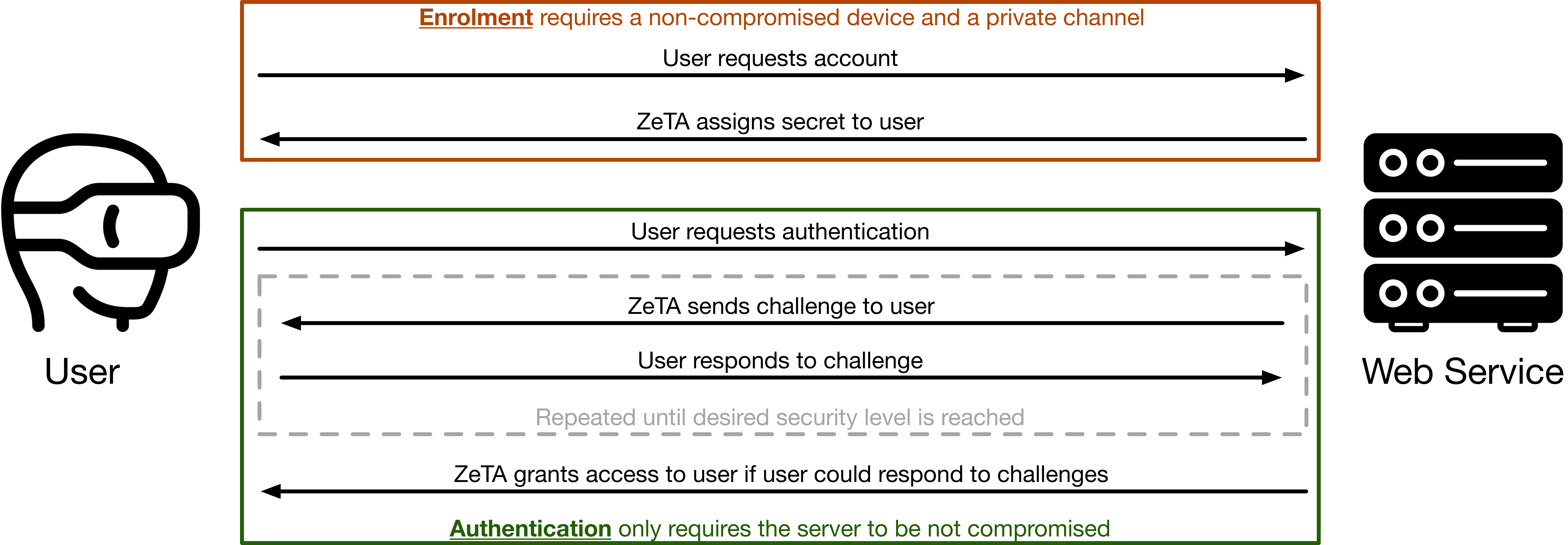}
    \caption{Overview of Zero-Trust Authentication (ZeTA) in AR \& VR Context.}
    \label{fig:zeta-procedure}
\end{figure*}

The proposed authentication scheme is based on our previous research: the Zero-Trust Authentication (ZeTA) protocol~\cite{gutmann2016zeta}. 
In this paper we describe how ZeTA can be applied to the AR \& VR context. 
Our future research goal is to implement the proposed authentication scheme using a user-centred development approach and conduct user studies to evaluate its usability and users' risk perception. Note, since organizations aim to provide their products and services worldwide, it is in particular interesting to understand the cultural differences in the use and perception of upcoming technologies like AR \& VR. 

The importance of social and cultural aspects when investigating the acceptability and appropriateness of technology are shown in many papers~\cite{benyon2005designing, kamppuri2006expanding, tractinsky1997aesthetics,dev2019personalized}. Hofstede’s~\cite{sondergaard2001culture} five cultural dimensions (namely power distance, individualism, masculinity, uncertainty avoidance and long-term orientation) are widely used to quantify national differences. These cultural dimensions showed many times an association towards technology use~\cite{van2003effect, erumban2006cross, al2002extending}. Some studies also discovered differences on perceived usability among different cultures \cite{noiwan2006cultural,reinecke2011improving}.

The impact of cultural aspects on the use and acceptance of HMDs and authentication schemes has yet to be determined. Thus, the study is going to be conducted in Germany and the U.S. for cross-cultural analysis.

\section{Related Work}

Prior research has proposed and developed different authentication schemes on HMDs. Yu et al.~\cite{yu2016exploration} and George et al.~\cite{george2017seamless} investigated well-established concepts for the VR context, such as PINs or 2D and 3D sliding patterns within VR environments. These concepts, though helpful for authentication, have some security concerns. For example, bystanders can observe or even record the movement which can help them to guess the password from the controller's action.

Additionally, for AR devices like Google Glass, Islam et al.~\cite{islam2018glasspass} proposed tapping gestures on the glasses' temple and use tapping patterns as a means to authenticate. Winkler et al.~\cite{Winkler:2015gs} introduced an authentication method that is more resistant to observations by using AR glasses in combination with the smartphone. The glasses show a randomly created PIN pad on the private display according to which the user can input password through their smartphone. Other proposals include biometric authentication based on head and body movement~\cite{mustafa2018unsure, miller2020within, li2016whose} or the human visual system~\cite{khamis2018vrpursuits, luo_oculock_2020, li2017accurate}. These proposals require either additional hardware (such as a smartphone) or a training phase to capture the user's biometric pattern. In contrast, our proposal requires neither.  

For any proposal aiming to advance authentication for AR \& VR devices, investigating societal and cultural aspects in technology adoption is critical. Prior studies have shown that authentication behaviour, usage, and experience is influenced vastly by age~\cite{das2019towards}, cultural differences~\cite{aljahdali2013affect}, and geographical locations~\cite{riley2009culture,volkamer2018replication,petrie2016cultural}. Riley et al. investigated regional differences in the perception of biometric authentication in India, South Africa, and the United Kingdom~\cite{riley2009culture}. Volkamer et al. observed in a field study PIN usage at ATMs and in various electronic payment scenarios in Germany, Sweden, and the United Kingdom~\cite{volkamer2018replication}. Given prior evidence, it is essential to evaluate the impact of different countries when designing a new authentication scheme, especially for new-age technologies. These technologies, such as AR \& VR are used worldwide where the demographical, societal, and cultural impact can play a critical role.

Yet, in the AR \& VR space, we found very little research on cross-cultural aspects. Jung et al.~\cite{jung2018cross} and Lee et al.~\cite{lee2015examining} explored the cultural differences in the adoption of mobile AR in South Korea and Ireland. Few studies investigated the effect of web-based AR on online shopping and compared results from different countries inside of Europe~\cite{pantano2017enhancing, gautier2016ar}. These studies identified differences in the use and perception of mobile and web-based AR applications between countries. Despite such critical research, to our knowledge, there are no cross-cultural studies in AR \& VR with HMDs. Thus, comparing HMD usage in different countries in the context of authentication will be novel and, therefore, very valuable. 

\section{Proposed Solution}
The goal of this work is to propose an authentication scheme for the AR \& VR devices, which is resistant to observation and only relies on the sensors integrated into the most AR \& VR HMDs. Our proposed authentication scheme is based on our previous research on observation resistant authentication: the Zero-Trust Authentication (ZeTA) protocol~\cite{gutmann2016zeta}. Here, we first provide a summary of the ZeTA protocol and explain how it could be applied in the AR \& VR context.

\subsection{Zero-Trust Authentication (ZeTA)}
\label{sec:zeta-description}
ZeTA is a knowledge-based authentication protocol, i.e., the user has to memorize a secret analogously to text passwords. In this section we describe its working principle, which is also illustrated in \cref{fig:zeta-procedure}.

The general idea of ZeTA is to expand upon the human capacity to build up semantic networks of related concepts and is thus based on innate human-based computation. To that end, ZeTA requires a knowledge base of concepts (e.g., words or symbols) and their semantic relations. The users' secrets in ZeTA consist of two or more concepts and logical connections between them (i.e., AND, OR, NOT), e.g., \lq\lq yellow OR wheel\rq\rq. This secret is generated and assigned to the user by ZeTA during the enrolment of the user. The enrolment has to be performed through a private channel between the system and the user. 

The authentication is based on a challenge-response interaction. The user has to determine whether a specific attribute is related to their secret or not, e.g., if the secret was \lq\lq yellow OR wheel\rq\rq~ and the challenge was \lq\lq sunflower\rq\rq, then the correct answer would be \lq\lq yes\rq\rq. 
Note that all challenges are pre-generated as part of the creation of the user secret and stored as described in~\cite{gutmann2016zeta}. Thereby, the secret is chosen such that it partitions the knowledge base equally in yes and no challenges (i.e., half of the attributes are related to the secret and half of the attributes are not related to the secret).

Due to its design, ZeTA can allow errors in responses by the users to compensate for innate differences in users' interpretations of the semantic relations between concepts. This can potentially increase ZeTAs usability but might impair security if the two are not carefully balanced. It also highlights the importance of cultural effects. The system repeats the challenge-response protocol until the desired certainty threshold is achieved; i.e., the probability of the user being an impostor is sufficiently small. Consequently, ZeTA can be scaled seamlessly to arbitrary security levels. When user errors are not allowed during an authentication attempt, according to~\cite{gutmann2016zeta} ZeTa can easily reach PIN-level security with 14 challenges. The usual online guessing threshold of $10^6$~\cite{Florencio:2014tu} can be achieved as easily using 25 challenges, while even allowing for one error by the user~\cite{gutmann2016zeta}.

As stated above, the enrolment procedure of ZeTA relies on a private channel. In contrast, after the enrolment, ZeTA was designed with the threat model as introduced by Matsumoto and Imai~\cite{Matsumoto:1991fi} in mind. The attacker can compromise the communication channels and even the user's device. Thus, ZeTA relies only on the server being secure. Proofs for lower bounds on the number of observations required to learn a secret based on a probably approximately correct learning model are presented in the original publication~\cite{gutmann2016zeta}.
 
\subsection{Application in the AR \& VR context}
\label{sec:arvr-zeta}
Augmented and Virtual Reality HMDs provide various interaction methods depending on the capabilities of the device. Examples of input systems are controller, head movement, gesture, and voice recognition. The core output system is the private display (i.e., optics that create the virtual image) combined with audio.  
The idea underlying the usage of ZeTA in the AR \& VR context is that the challenge is shown on the display of the HMDs. The user responses are entered using input options, which can be found in most of the AR \& VR HMDs. Thus, we avoid dependencies on additional hardware. Concerning the entry of the response to the system's challenge by the user, the following interaction options can be used: 1) voice control, 2) head movement, and 3) buttons on the VR controller or touch controls on the AR glasses. Additionally, finding the right number of challenges as a trade-off between usability and security while considering the specifics of the AR/VR context is an important aspect of the development of the ZeTA implementations for our user study.

The advantage of using ZeTA as an authentication scheme for the AR/VR context is that shoulder-surfing resistance does not need to be empirically evaluated due to the aforementioned security proofs. Therefore, the lower bound of needed observations holds no matter whether the attacker observes the communication channel, the user's interaction, or even the private display of the HMD. The user input can even be processed by the web server and does not need to burden the capacity of the HMD without impairing the user's privacy. The only time the user is required to use a non-compromised device and a private channel is when being assigned the secret by ZeTA during enrolment. 

\subsection{Future Design and Implementation}
The proposed authentication scheme will be implemented as mock-up for both AR \& VR HMDs, as well as for each interaction method. The development is based on a human-centered design approach: the mock-ups are tested and improved iteratively by evaluating different design variations of the outputs and inputs with users to maximize the authentication scheme's usability. Options for the output to show the challenges are text, image, and audio. Options for the input of the responses are: voice, head movement, and buttons/touch controls (cf.~\cref{sec:arvr-zeta}). There might also be different approaches to give feedback to the user after answering each challenge or to proceed from one challenge to the next one.

\section{Proposed Methodology for User Evaluation}
\begin{figure*}[htp!]
    \centering
    \includegraphics[width=0.8\linewidth]{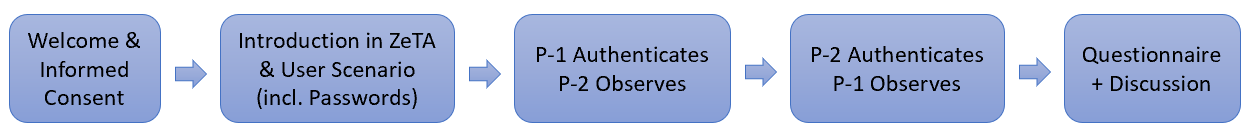}
    \caption{Main study protocol.}
    \label{fig:mainstudy}
\end{figure*}

As future work, we will evaluate the three interaction methods of the proposed authentication scheme through in-lab user studies. We are planning to use Google Glass for the AR application and the Oculus Rift S for the VR application. The study design is built upon our research on shoulder-surfing resistant authentication using gamepads~\cite{mayer2019don}.

\subsection{Research Goal}
The evaluation of the authentication scheme for each of the AR \& VR HMDs and each of the three interaction methods (voice, head movement, and touch/press) will be based on usability criteria and users' risk perception regarding the authentication protocol security. Usability is measured by users effectiveness, efficiency and satisfaction with the authentication scheme. 
Thus, our research goal for future work is: \\
\textit{Identifying the best interaction method for authenticating through ZeTA on both, AR \& VR HMDs, i.e., the method that provides the highest effectiveness, efficiency, and satisfaction as well as the lowest perceived risk by users regarding the security of the authentication process.}\\
We aim to inspect the cross-cultural influence by conducting identical studies in Germany and the United States. Germany and the U.S. are interesting cultures to compare because of their global influence in the field of technology~\cite{greenstein2008comparison,morgan2004360}. Both of these nations share much in common (democratic governments, similar linguistic roots), they also have some interesting differences (ethical heterogeneity, capitalistic versus socialistic approach)~\cite{schmuck2000intrinsic}. Additionally, it is predicted that the AR/VR market will rise globally, especially in U.S. (96.1\% Compound annual growth rate (CAGR))  and in Western European countries (104.2\% CAGR), including Germany~\cite{IDC2019}.

\subsection{Study Protocol}
\begin{figure}[b!]
    \centering
    \includegraphics[width=0.8\linewidth]{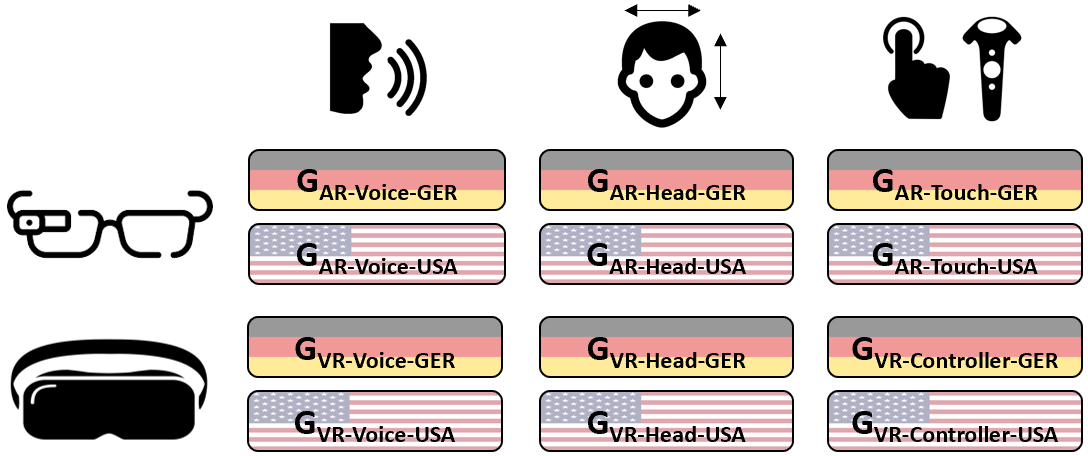}
    \caption{Allocation of the groups in the main user study.}
    \label{fig:usergroups}
\end{figure}
After completing the implementation, a pre-study is planned to pilot and refine the study protocol of the main study, which is described below. The authentication scheme is tested with each combination of the device (i.e., AR, VR) and interaction method (i.e., voice, head gestures, button/touch controls) regarding its usability and users' risk perception regarding its security mechanism. The study will be conducted in both, Germany and the United States. Therefore, 12 (2x3x2) groups are used to collect data as visualized in~\cref{fig:usergroups}. 

Each participant will test all three interaction methods. To avoid first-order carryover effects, the allocation of the participants will be specified with the Latin Square Design~\cite{coleman2018designing} that counterbalances sequential effects. 
The procedure of the main study is presented in figure \ref{fig:mainstudy}. We will ask two participants to come to the lab simultaneously. Both of the participants will receive an explanation of the ZeTA scheme and will be given a user scenario with three different randomly generated passwords. Then, we will run a 3-step evaluation process: 
\begin{enumerate}
    \item Participant-1 authenticates on the HMD three times. Participant-2 observes the process. 
    \item Now they change roles: participant-2 authenticates on the HMD three times. Participant-1 observes the process. 
    \item Both participants answer questions in a survey as well as we conclude with a short semi-structured interview.
\end{enumerate}

By having two participants in the lab simultaneously, we aim to create a higher validity setting with respect to evaluating users' risk perception. Secrets will be assigned to the participants by the system. Each of them will have time alone to memorize their secret. As a baseline for the configuration, we propose to use the online guessing resistance threshold of $10^6$~\cite{Florencio:2014tu}. This is in line with the envisioned types of accounts used on the HMDs (e.g., purchasing media content from on online service). Before conducting the study, we will ask for ethical approval. Participants will be compensated based on the minimum wage regulations in the U.S. and Germany. 

The effectiveness will be measured by the ratio of correct password entries among the three. Efficiency will be assessed by the average time needed for authentication across the three passwords. Satisfaction will be measured with the System Usability Scale (SUS) that covers users’ subjective reactions to using the scheme~\cite{brooke1996sus}. To examine the user's risk perception, the scales proposed by Fischhoff et al.~\cite{fischhoff1978safe}, Liang \& Xue~\cite{liang2010understanding}, and Das~\cite{das2020risk} will be adapted to our use case. The risk perception metric is defined by nine characteristics of the risk: 1) voluntariness, 2) immediacy, 3) knowledge of the exposed, 4) knowledge of experts, 5) control, 6) newness, 7) common-dread, 8) chronic-catastrophic, and 9) severity. Offline, this framework informed four decades of research in risk perception and public policy in a diversity of risk domains, e.g., environmental risk~\cite{flynn1994gender} and health risk~\cite{johnson1995presenting}. Online, this framework has been used to explain perceptions of technical security risks~\cite{camp2006mental,das2020user} and insider threats~\cite{farahmand2013understanding}. 

\section*{Acknowledgments}

This work was supported by the German Federal Ministry of Education and Research (BMBF) in the Competence Center for Applied Security Technology (KASTEL), Karlsruhe Institute of Technology; Secure and Privacy Research in New-Age Technology (SPRINT) Lab, University of Denver; and Human and Technical Security (HATS) Lab, Indiana University. Any opinions, findings, and conclusions or recommendations expressed in this material are solely those of the author(s). 

\bibliographystyle{plain}

\bibliography{ARXIV_2020_ARVR_WAYS.bib}

\begin{thebibliography}{10}

\bibitem{al2002extending}
Said~S Al-Gahtani.
\newblock Extending the technology acceptance model beyond its country of
  origin: a cultural test in western europe.
\newblock In {\em Advanced Topics in Information Resources Management, Volume
  1}, pages 158--183. IGI Global, 2002.

\bibitem{aljahdali2013affect}
Hani~Moaiteq Aljahdali and Ron Poet.
\newblock The affect of familiarity on the usability of recognition-based
  graphical passwords: Cross cultural study between saudi arabia and the united
  kingdom.
\newblock In {\em 12th IEEE TrustCom}, pages 1528--1534, 2013.

\bibitem{benyon2005designing}
David Benyon, Phil Turner, and Susan Turner.
\newblock {\em Designing interactive systems: People, activities, contexts,
  technologies}.
\newblock Pearson Education, 2005.

\bibitem{brooke1996sus}
John Brooke.
\newblock Sus: a “quick and dirty'usability.
\newblock {\em Usability evaluation in industry}, page 189, 1996.

\bibitem{camp2006mental}
L.~Jean Camp.
\newblock Mental models of privacy and security.
\newblock {\em SSRN Electronic Journal}, 2006.

\bibitem{chan2015glass}
Pan Chan, Tzipora Halevi, and Nasir Memon.
\newblock Glass otp: Secure and convenient user authentication on google glass.
\newblock In {\em Financial Cryptography and Data Security}, pages 298--308.
  Springer, 2015.

\bibitem{coleman2018designing}
Renita Coleman.
\newblock {\em Designing experiments for the social sciences: How to plan,
  create, and execute research using experiments}.
\newblock Sage publications, 2018.

\bibitem{IDC2019}
IDC: International~Data Corporation.
\newblock Worldwide spending on augmented and virtual reality expected to reach
  \$18.8 billion in 2020, according to idc.
\newblock \url {https://www.idc.com/getdoc.jsp?containerId=prUS45679219}, 2019.
\newblock Accessed: 2020-07-17.

\bibitem{das2020risk}
Sanchari Das.
\newblock {\em A Risk-reduction-based Incentivization Model for Human-centered
  Multi-factor Authentication}.
\newblock PhD thesis, Indiana University, 2020.

\bibitem{das2020user}
Sanchari Das, Jacob Abbott, Shakthidhar Gopavaram, Jim Blythe, and L~Jean Camp.
\newblock User-centered risk communication for safer browsing.
\newblock In {\em Proceedings of the First Asia USEC-Workshop on Usable
  Security, In Conjunction with the Twenty-Fourth International Conference
  International Conference on Financial Cryptography and Data Security}, 2020.

\bibitem{das2019towards}
Sanchari Das, Andrew Kim, Ben Jelen, Joshua Streiff, L~Jean Camp, and Lesa
  Huber.
\newblock Towards implementing inclusive authentication technologies for older
  adults.
\newblock {\em Who Are You}, 2019.

\bibitem{dev2019personalized}
Jayati Dev, Sanchari Das, Yasmeen Rashidi, and L~Jean Camp.
\newblock Personalized whatsapp privacy: Demographic and cultural influences on
  indian and saudi users.
\newblock {\em Available at SSRN 3391021}, 2019.

\bibitem{erumban2006cross}
Abdul~Azeez Erumban and Simon~B De~Jong.
\newblock Cross-country differences in ict adoption: A consequence of culture?
\newblock {\em Journal of world business}, 41(4):302--314, 2006.

\bibitem{farahmand2013understanding}
Fariborz Farahmand and Eugene~H Spafford.
\newblock Understanding insiders: An analysis of risk-taking behavior.
\newblock {\em Information Systems Frontiers}, 15(1):5--15, 2013.

\bibitem{fischhoff1978safe}
Baruch Fischhoff, Paul Slovic, Sarah Lichtenstein, Stephen Read, and Barbara
  Combs.
\newblock How safe is safe enough? a psychometric study of attitudes towards
  technological risks and benefits.
\newblock {\em Policy sciences}, 9(2):127--152, 1978.

\bibitem{Florencio:2014tu}
Dinei Flor{\^e}ncio, Cormac Herley, and Paul~C van Oorschot.
\newblock {An Administrator{\textquoteright}s Guide to Internet Password
  Research}.
\newblock In {\em Large Installation System Administration Conference}, pages
  35--52. USENIX, 2014.

\bibitem{flynn1994gender}
James Flynn, Paul Slovic, and Chris~K Mertz.
\newblock Gender, race, and perception of environmental health risks.
\newblock {\em Risk analysis}, 14(6):1101--1108, 1994.

\bibitem{gafurov2006biometric}
Davrondzhon Gafurov, Kirsi Helkala, and Torkjel S{\o}ndrol.
\newblock Biometric gait authentication using accelerometer sensor.
\newblock {\em JCP}, 1(7):51--59, 2006.

\bibitem{gautier2016ar}
St{\'e}phanie Gautier, Claire Gauzente, and Maiju Aikala.
\newblock Are ar shopping services valued the same way throughout europe? a
  four-country q-investigation.
\newblock {\em Systemes d'information management}, 21(1):69--102, 2016.

\bibitem{george2017seamless}
Ceenu George, Mohamed Khamis, Emanuel von Zezschwitz, Marinus Burger, Henri
  Schmidt, Florian Alt, and Heinrich Hussmann.
\newblock Seamless and secure vr: Adapting and evaluating established
  authentication systems for virtual reality.
\newblock In {\em NDSS}, 2017.

\bibitem{greenstein2008comparison}
Marilyn Greenstein-Prosch, Thomas~E McKee, and Reiner Quick.
\newblock A comparison of the information technology knowledge of united states
  and german auditors.
\newblock {\em The International Journal of Digital Accounting Research},
  8(14):45--79, 2008.

\bibitem{gugenheimer2019challenges}
Jan Gugenheimer, Christian Mai, Mark McGill, Julie Williamson, Frank Steinicke,
  and Ken Perlin.
\newblock Challenges using head-mounted displays in shared and social spaces.
\newblock In {\em EA CHI}, pages 1--8, 2019.

\bibitem{gutmann2016zeta}
Andreas Gutmann, Karen Renaud, Joseph Maguire, Peter Mayer, Melanie Volkamer,
  Kanta Matsuura, and J{\"o}rn M{\"u}ller-Quade.
\newblock Zeta-zero-trust authentication: Relying on innate human ability, not
  technology.
\newblock In {\em EuroS\&P}, pages 357--371. IEEE, 2016.

\bibitem{happa2019cyber}
Jassim Happa, Mashhuda Glencross, and Anthony Steed.
\newblock Cyber security threats and challenges in collaborative mixed-reality.
\newblock {\em Frontiers in ICT}, 6:5, 2019.

\bibitem{islam2018glasspass}
MD~Rasel Islam, Doyoung Lee, Liza~Suraiya Jahan, and Ian Oakley.
\newblock Glasspass: Tapping gestures to unlock smart glasses.
\newblock In {\em AH}, pages 1--8, 2018.

\bibitem{johnson1995presenting}
Branden~B Johnson and Paul Slovic.
\newblock Presenting uncertainty in health risk assessment: initial studies of
  its effects on risk perception and trust.
\newblock {\em Risk Analysis}, 15(4):485--494, 1995.

\bibitem{jung2018cross}
Timothy~Hyungsoo Jung, Hyunae Lee, Namho Chung, and M~Claudia tom Dieck.
\newblock Cross-cultural differences in adopting mobile augmented reality at
  cultural heritage tourism sites.
\newblock {\em IJCHM}, 2018.

\bibitem{kamppuri2006expanding}
Minna Kamppuri, Roman Bednarik, and Markku Tukiainen.
\newblock The expanding focus of hci: case culture.
\newblock In {\em Proceedings of the 4th Nordic conference on Human-computer
  interaction: changing roles}, pages 405--408, 2006.

\bibitem{khamis2018vrpursuits}
Mohamed Khamis, Carl Oechsner, Florian Alt, and Andreas Bulling.
\newblock Vr pursuits: interaction in virtual reality using smooth pursuit eye
  movements.
\newblock In {\em AVI}, pages 1--8, 2018.

\bibitem{lee2015examining}
Hyunae Lee, Namho Chung, and Timothy Jung.
\newblock Examining the cultural differences in acceptance of mobile augmented
  reality: Comparison of south korea and ireland.
\newblock In {\em Information and communication technologies in tourism}, pages
  477--491. Springer, 2015.

\bibitem{li2016whose}
Sugang Li, Ashwin Ashok, Yanyong Zhang, Chenren Xu, Janne Lindqvist, and Macro
  Gruteser.
\newblock Whose move is it anyway? authenticating smart wearable devices using
  unique head movement patterns.
\newblock In {\em PerCom}, pages 1--9. IEEE, 2016.

\bibitem{li2017accurate}
Yung-Hui Li and Po-Jen Huang.
\newblock An accurate and efficient user authentication mechanism on smart
  glasses based on iris recognition.
\newblock {\em Mobile Information Systems}, 2017.

\bibitem{liang2010understanding}
Huigang Liang, Yajiong~Lucky Xue, et~al.
\newblock Understanding security behaviors in personal computer usage: A threat
  avoidance perspective.
\newblock {\em Journal of the association for information systems}, 11(7):1,
  2010.

\bibitem{luo_oculock_2020}
Shiqing Luo, Anh Nguyen, Chen Song, Feng Lin, Wenyao Xu, and Zhisheng Yan.
\newblock {OcuLock}: Exploring human visual system for authentication in
  virtual reality head-mounted display.
\newblock In {\em NDSS}, 2020.

\bibitem{Matsumoto:1991fi}
Tsutomu Matsumoto and Hideki Imai.
\newblock {Human Identification Through Insecure Channel}.
\newblock In {\em EUROCRYPT}, pages 409--421. Springer, 1991.

\bibitem{mayer2019don}
Peter Mayer, Nina Gerber, Benjamin Reinheimer, Philipp Rack, Kristoffer Braun,
  and Melanie Volkamer.
\newblock I (don't) see what you typed there! shoulder-surfing resistant
  password entry on gamepads.
\newblock In {\em CHI}, pages 1--12, 2019.

\bibitem{miller2020within}
Robert Miller, Natasha~Kholgade Banerjee, and Sean Banerjee.
\newblock Within-system and cross-system behavior-based biometric
  authentication in virtual reality.
\newblock In {\em IEEE VR}, pages 311--316, 2020.

\bibitem{morgan2004360}
Michael Morgan and Laura Borns.
\newblock 360 degrees of usability.
\newblock In {\em CHI'04 Extended Abstracts on Human Factors in Computing
  Systems}, pages 795--809, 2004.

\bibitem{mustafa2018unsure}
Tahrima Mustafa, Richard Matovu, Abdul Serwadda, and Nicholas Muirhead.
\newblock Unsure how to authenticate on your vr headset? come on, use your
  head!
\newblock In {\em ACM IWSPA}, pages 23--30, 2018.

\bibitem{noiwan2006cultural}
Jantawan Noiwan and Anthony~F Norcio.
\newblock Cultural differences on attention and perceived usability:
  Investigating color combinations of animated graphics.
\newblock {\em International journal of Human-computer studies},
  64(2):103--122, 2006.

\bibitem{pantano2017enhancing}
Eleonora Pantano, Alexandra Rese, and Daniel Baier.
\newblock Enhancing the online decision-making process by using augmented
  reality: A two country comparison of youth markets.
\newblock {\em Journal of Retailing and Consumer Services}, 38:81--95, 2017.

\bibitem{petrie2016cultural}
Helen Petrie and Burak Merdenyan.
\newblock Cultural and gender differences in password behaviors: Evidence from
  china, turkey and the uk.
\newblock In {\em Nordic Conference on CHI}, pages 1--10, 2016.

\bibitem{reinecke2011improving}
Katharina Reinecke and Abraham Bernstein.
\newblock Improving performance, perceived usability, and aesthetics with
  culturally adaptive user interfaces.
\newblock {\em ACM TOCHI}, 18(2):1--29, 2011.

\bibitem{ABI2017}
ABI Research and Qualcomm.
\newblock Augmented and virtual reality: the first wave of 5g killer apps.
\newblock \url
  {https://www.qualcomm.com/news/onq/2017/02/01/vr-and-ar-are-pushing-limits-connectivity-5
  g-our-rescue}, 2017.
\newblock Accessed: 2020-06-17.

\bibitem{riley2009culture}
Chris Riley, Kathy Buckner, Graham Johnson, and David Benyon.
\newblock Culture \& biometrics: regional differences in the perception of
  biometric authentication technologies.
\newblock {\em AI \& society}, 24(3):295--306, 2009.

\bibitem{roberts2014visualization}
Jonathan~C Roberts, Panagiotis~D Ritsos, Sriram~Karthik Badam, Dominique
  Brodbeck, Jessie Kennedy, and Niklas Elmqvist.
\newblock Visualization beyond the desktop--the next big thing.
\newblock {\em IEEE Computer Graphics and Applications}, 34(6):26--34, 2014.

\bibitem{schmuck2000intrinsic}
Peter Schmuck, Tim Kasser, and Richard~M Ryan.
\newblock Intrinsic and extrinsic goals: Their structure and relationship to
  well-being in german and us college students.
\newblock {\em Social Indicators Research}, 50(2):225--241, 2000.

\bibitem{sondergaard2001culture}
Mikael S{\o}ndergaard and Geert Hofstede.
\newblock Culture's consequences: comparing values, behaviours, institutions,
  and organizations across nations.
\newblock {\em International Journal of Cross Cultural Management}, pages
  243--246, 2001.

\bibitem{tractinsky1997aesthetics}
Noam Tractinsky.
\newblock Aesthetics and apparent usability: empirically assessing cultural and
  methodological issues.
\newblock In {\em Proceedings of the ACM SIGCHI Conference on Human factors in
  computing systems}, pages 115--122, 1997.

\bibitem{van2003effect}
Yvonne~M Van~Everdingen and Eric Waarts.
\newblock The effect of national culture on the adoption of innovations.
\newblock {\em Marketing letters}, 14(3):217--232, 2003.

\bibitem{volkamer2018replication}
Melanie Volkamer, Andreas Gutmann, Karen Renaud, Paul Gerber, and Peter Mayer.
\newblock Replication study: A cross-country field observation study of real
  world pin usage at atms and in various electronic payment scenarios.
\newblock In {\em SOUPS}, pages 1--11, 2018.

\bibitem{Winkler:2015gs}
Christian Winkler, Jan Gugenheimer, Alexander De~Luca, Gabriel Haas, Philipp
  Speidel, David Dobbelstein, and Enrico Rukzio.
\newblock {Glass Unlock: Enhancing Security of Smartphone Unlocking through
  Leveraging a Private Near-eye Display}.
\newblock In {\em CHI}, pages 1407--1410. ACM, 2015.

\bibitem{yu2016exploration}
Zhen Yu, Hai-Ning Liang, Charles Fleming, and Ka~Lok Man.
\newblock An exploration of usable authentication mechanisms for virtual
  reality systems.
\newblock In {\em APCCAS}, pages 458--460. IEEE, 2016.

\end{thebibliography}

\end{document}